\documentclass[aps,prb,twocolumn,superscriptaddress,showkeys,floatfix]{revtex4}

\usepackage{array}
\usepackage{graphicx,color}
\usepackage{multirow,slashbox}
\usepackage{textcomp}
\usepackage{amsmath}
\usepackage{subeqnarray}
\usepackage{cases}

\graphicspath{{figs/}}
\begin{document}

\title{Anomalous strain-dependent thermal conductivity in superelastic screw-dislocated graphites}

\author{Yu Li}
    \affiliation{Shanghai Key Laboratory of Mechanics in Energy Engineering, Shanghai Institute of Applied Mathematics and Mechanics, Shanghai Frontier Science Center of Mechanoinformatics, School of Mechanics and Engineering Science, Shanghai University, Shanghai 200072, People\textquotesingle s Republic of China}
    \affiliation{Institute of High Performance Computing (IHPC), Agency for Science, Technology and Research (A*STAR), 1 Fusionopolis Way, \#16-16 Connexis, Singapore 138632, Republic of Singapore}

\author{Zhiqiang Zhao}
   \affiliation{Institute of High Performance Computing (IHPC), Agency for Science, Technology and Research (A*STAR), 1 Fusionopolis Way, \#16-16 Connexis, Singapore 138632, Republic of Singapore}
    \affiliation{State Key Laboratory of Mechanics and Control for Aerospace Structures, Key Laboratory for Intelligent Nano Materials and Devices of Ministry of Education, and Institute for Frontier Science, Nanjing University of Aeronautics and Astronautics, Nanjing 210016, China}

\author{Zhuhua Zhang}
    \affiliation{State Key Laboratory of Mechanics and Control for Aerospace Structures, Key Laboratory for Intelligent Nano Materials and Devices of Ministry of Education, and Institute for Frontier Science, Nanjing University of Aeronautics and Astronautics, Nanjing 210016, China}
 
\author{Yong-Wei Zhang}
    \altaffiliation{Corresponding author: zhangyw@ihpc.a-star.edu.sg (Y.-W. Zhang)}
    \affiliation{Institute of High Performance Computing (IHPC), Agency for Science, Technology and Research (A*STAR), 1 Fusionopolis Way, \#16-16 Connexis, Singapore 138632, Republic of Singapore}
   
\author{Jin-Wu Jiang}
    \altaffiliation{Corresponding author: jwjiang5918@hotmail.com (J.-W.Jiang)}
    \affiliation{Shanghai Key Laboratory of Mechanics in Energy Engineering, Shanghai Institute of Applied Mathematics and Mechanics, Shanghai Frontier Science Center of Mechanoinformatics, School of Mechanics and Engineering Science, Shanghai University, Shanghai 200072, People\textquotesingle s Republic of China}

\date{\today}
\begin{abstract}
The design of strain-stable, or even strain-enhanced thermal transport materials is critical for stable operation of high-performance electronic devices. However, most nanomaterials suffer from strain-induced degradation, with even minor tensile strains markedly reducing thermal conductivity. Here, we demonstrate that screw-dislocated graphites (SDGs), recently identified as topological semimetals, display an unusual increase in cross-plane thermal conductivity under both tensile and compressive strains, revealed by high-accuracy machine-learning-potential-driven non-equilibrium molecular dynamics. Notably, SDGs exhibit over 100\% enhancement under tensile strains up to 80\% along the dislocation axis, arising from strain-induced increase in dislocation interface tilt angle that elongates the effective heat transfer paths. Their thermal conductivity surpasses multilayer graphene by an order of magnitude. An analytical model is further derived linking thermal conductivity to dislocation number and strain, offering a predictive framework for designing strain-tunable screw-dislocated structures. These findings highlight SDGs as a promising platform for high-performance electronic and wearable devices with tunable thermal properties.
\end{abstract}


\maketitle
\pagebreak

\section{Introduction}

\begin{figure*}[htpb]
  \begin{center}
    \scalebox{1}[1]{\includegraphics[width=15cm]{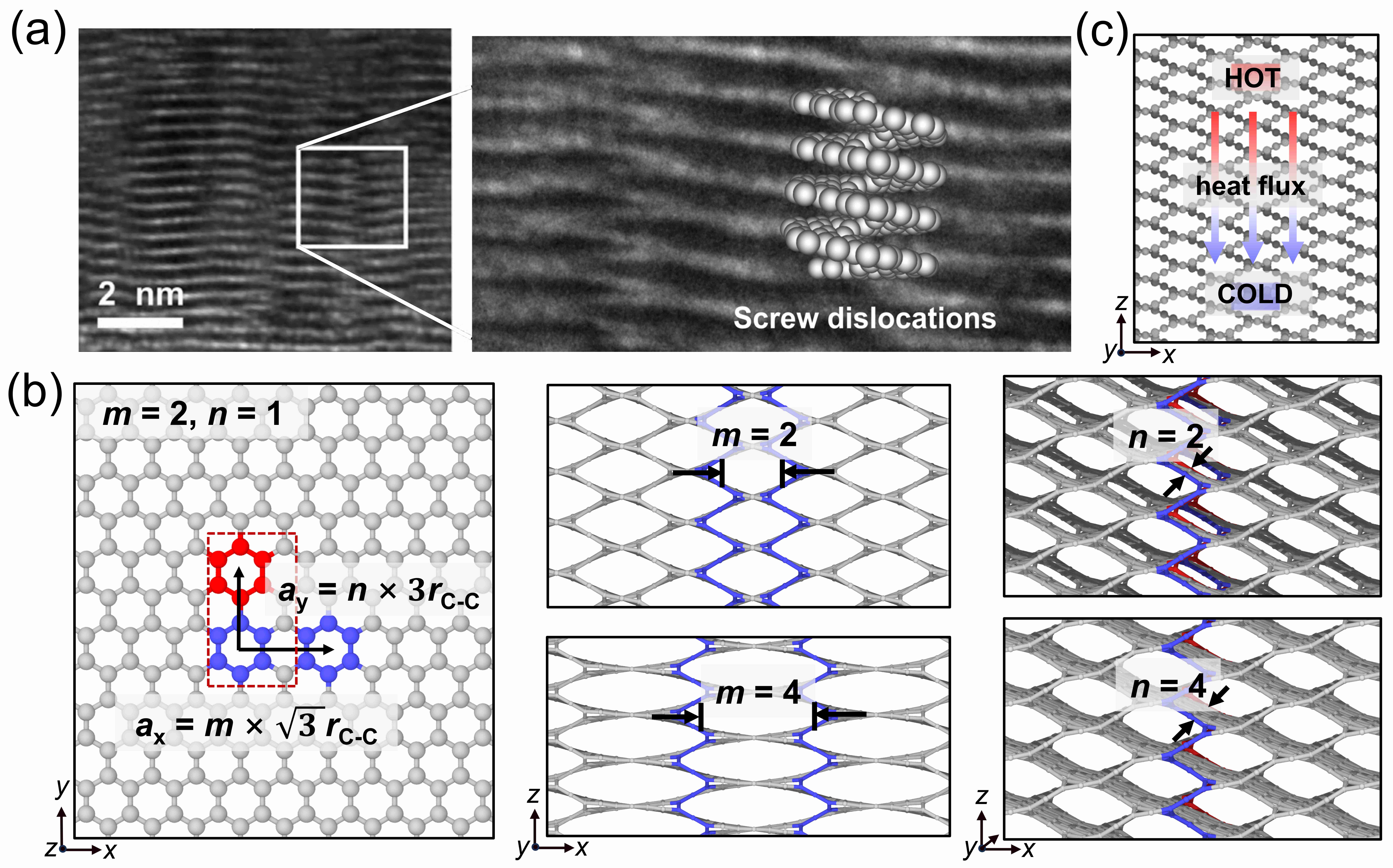}}
  \end{center}
  \caption{(Color online) Structures of screw-dislocated graphites (SDGs). (a) The high-resolution transmission electron microscopy images of screw dislocations in graphite, reproduced from Ref. 25, ©~2023~The Authors. Published by Elsevier Ltd., distributed under the terms of the Creative Commons CC-BY license. (b) Schematics of SDGs, featuring paired dislocations with the same chirality (marked in blue) along the $x$-direction and paired dislocations with complementary chirality (marked in red) along the $y$-direction in graphene. The smallest repeat unit is given as marked by red dashed lines (left). Structures with different dislocation spacings along the $x$- (middle) and $y$-directions (right). (c) Set-up for the NEMD method. Heat flux flows from the hot region to the cold region. }
  \label{fig_model}
\end{figure*}

Low-dimensional topological carbon materials, exemplified by graphene, have provided ideal platforms for probing quantum phenomena owing to their massless Dirac fermions, topologically protected edge states, and ultrahigh carrier mobility.\cite{pesin2012spintronics, okuyama2019topological} Inspired by these systems, three-dimensional (3D) topological carbon allotropes, including bco-C16\cite{wang2016body} and bct-C16,\cite{cheng2017body} have subsequently been proposed by employing such low-dimensional structures as fundamental building units.\cite{wang2013new, wang2014new} These carbon allotropes host diverse topological states, including topological nodal-line,\cite{wang2016body, chen2022second} nodal-loop\cite{zhao2019topological, gao2018class} and Weyl semimetals,\cite{zhong2016towards, zhang2020symmetry} offering tremendous promise for low-dissipation transport and spintronics applications. While their exotic electronic features are well recognized, the practical integration of such materials into future high-performance devices will also hinge on their thermal properties. In particular, the increasing power density of modern electronics imposes stringent requirements on thermal management, where high thermal conductivity is critical for both heat dissipation and structural stability.\cite{balandin2008superior, subramaniam2014carbon, xin2015highly, guo2025generative}

In flexible electronic devices and micro/nano chip packaging, materials are inevitably subjected to bending, tensile, or compressive strains.\cite{yu2008raman, ha2016unconventional} Such mechanical strains can alter the crystal structure and thereby strongly modify transport properties. Due to distorted atomic networks, the structural stability and transport characteristics of 3D topological carbon phases are usually highly sensitive to external strain.\cite{gao2018class, zhong2017three} Moreover, most reported thermal transport materials exhibit a universal trend: tensile strain typically reduces the thermal conductivity owing to the reduced interatomic interactions, softened phonon modes, and enhanced three-phonon scattering; whereas compressive strain enhances it through strengthened interatomic interactions.\cite{xu1991theory, mounet2005first, parrish2014origins, he2019orbitally} Therefore, exploring and designing materials that combine stable topological electronic states with strain-stable, or even strain-enhanced, thermal transport is of critical importance for advancing the performance of next-generation electronic devices.

Screw-dislocated graphites (SDGs) have recently emerged as a promising candidate.\cite{zhao2019topological, zhao2020family} Their basic architecture incorporates two screw dislocations with same chirality along the zigzag direction and a pair with complementary chirality along the armchair directions of graphite fragments. As such, the long-range stress fields derived from dislocations are largely cancelled, thereby achieving excellent mechanical stability. By tuning dislocation spacing, a family of superelastic SDGs with novel topological semimetal states can be realized, which notably preserve their topological states under elastic strains up to 75\%. This extraordinary superelasticity not only prevents bond stretching that usually diminishes thermal transport, but also opens routes for wide-range modulation of thermal transport through strain engineering. Moreover, recent experimental breakthroughs further support the synthesis and future device application of such materials,\cite{wang2022synthesis} motivating systematic investigation of their strain-dependent thermal transport properties. 

In this work, we use non-equilibrium molecular dynamics (NEMD) simulations to investigate the cross-plane thermal conductivity along dislocation axis in SDGs and uncover anomalous strain-dependent thermal transport behavior. The introduction of screw dislocations enables the formation of interlayer covalent bonds, boosting the cross-plane thermal conductivity by more than an order of magnitude compared to conventional multilayer graphene (MLG). Strikingly, SDGs exhibit an anomalous enhancement of cross-plane thermal conductivity, increasing by more than 100\% under elastic tensile strains up to 80\% along the dislocation axis.  The counterintuitive enhancement arises from strain-induced increases in the dislocation interface tilt angle, which extensively extends heat current transport path. Meanwhile, the cross-plane thermal conductivity of SDGs increases by more than 700\% under elastic compressive strains up to 31\% owing to reinforced van der Waals interactions. Moreover, varying the number of dislocations generates distinct SDGs configurations, which significantly influence thermal transport properties. Accordingly, we derive an analytical expression that captures the dependence of cross-plane thermal conductivity on the dislocation numbers and applied strain. This provides fundamental insights for the rational design of screw-dislocated carbon structures with either high or low thermal conductivity tailored to practical application requirements. 

\section{Results and discussion}
\subsection{Structure of SDGs}

\begin{figure*}[tb]
  \begin{center}
    \scalebox{1.0}[1.0]{\includegraphics[width=16cm]{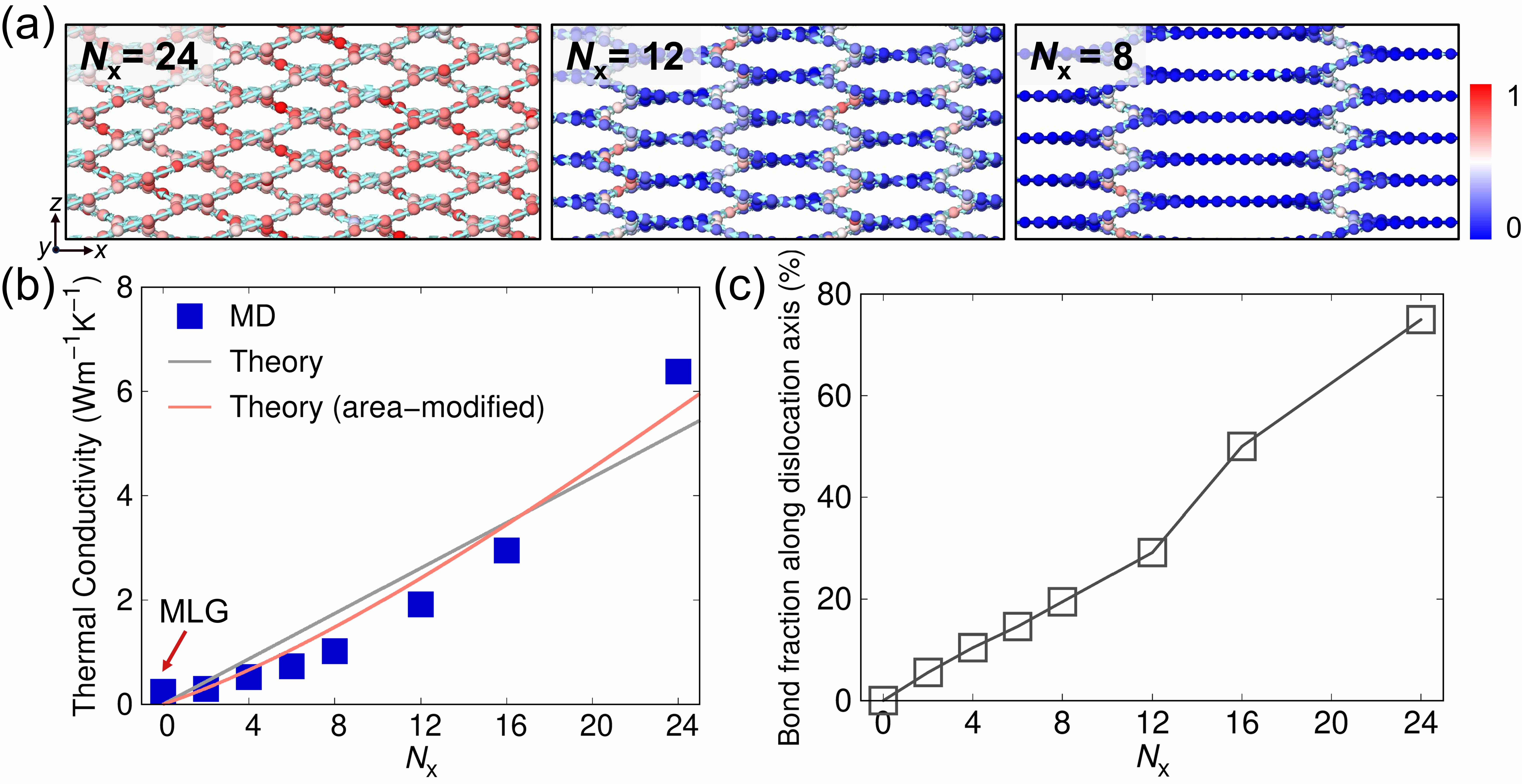}}
  \end{center}
  \caption{(Color online) Effect of screw dislocation number along $x$-direction $N_{\rm x}$ on the cross-plane thermal conductivity of SDGs. (a) Comparison of per-atom heat current distributions in SDGs with different $N_{\rm x}$, and only a small portion of the system (24576 atoms in total) is shown for clarity. Atom colors indicate the normalized magnitude of the heat current along the transport direction ($z$-direction) . Arrows represent both the magnitude and direction of the per-atom heat current. (b) Thermal conductivity and (c) fraction of covalent bonds along the dislocation axis as functions of dislocation number $N_{\rm x}$. The gray and red line are the theoretical results from equations (2) and (3), respectively.} 
  \label{fig_m_effect_TC}
\end{figure*}

 Fig.~\ref{fig_model}~(a) presents the microscopic image of carbon structure with screw dislocations in the experiment, where these dislocations wind through the graphene layers like a spiral staircase.\cite{martin2023graphite} The stable 3D carbon nanostructure studied in this work is constructed from graphene layers interconnected by paired screw dislocations with the same chirality along the zigzag direction ($x$-direction) and paired screw dislocations with complementary chirality along the armchair direction ($y$-direction) as illustrated in Fig.~\ref{fig_model}~(b). Adjacent screw dislocations are connected along the $x$-direction via van der Waals-stacked MLG regions, while along the $y$-direction, they are bridged by covalently bonded graphene ramps. Screw dislocations with the same chirality are periodically introduced along the $x$-direction with a spacing of $a_{\rm x}=m\times \sqrt 3 r_{\rm C-C}$, where $r_{\rm C-C} \approx 1.42$~{\text{\AA}} is the graphene bond length. Along the $y$-direction, dislocations with complementary chirality are introduced with a spacing of $a_{\rm y}=n\times 3 r_{\rm C-C}$. The dimensionless quantities $m$ and $n$ are the dislocation spacing parameter. The variables $N_{\rm x}$ and $N_{\rm y}$ are defined as the dislocation numbers along the $x$- and $y$-directions, respectively. A diverse set of SDGs can be generated by varying the dislocation numbers $N_{\rm x}$ and $N_{\rm y}$, while keeping the structural dimensions constant. 

\subsection{Effect of screw dislocation number on thermal conductivity}
 
\subsubsection{Dislocation number $N_{\rm x}$ effect}

\begin{figure*}[tb]
  \begin{center}
    \scalebox{1.0}[1.0]{\includegraphics[width=15cm]{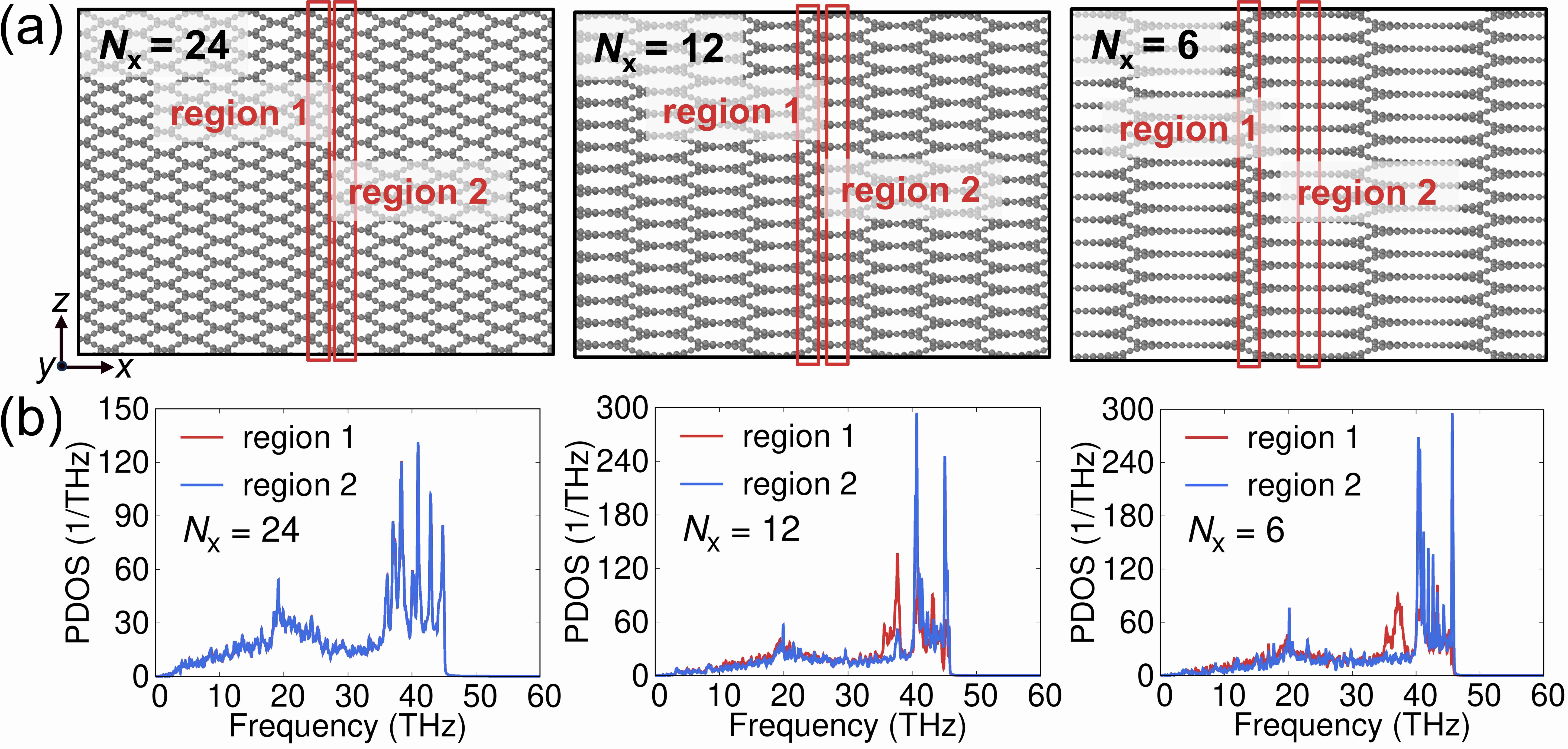}}
  \end{center}
  \caption{(Color online) Phonon density of states calculation. (a) Schematic of the partitioning of SDGs with different screw dislocation numbers $N_{\rm x}$. (b) Vibrational density of states of carbon atoms in regions 1 and 2 for SDGs with different $N_{\rm x}$.} 
  \label{fig_pdos_m_effect_TC}
\end{figure*}

We first investigate the influence of screw dislocation number on the thermal conductivity of SDGs. Fig.~\ref{fig_m_effect_TC}~(a) presents a representative segment of SDGs with varying numbers of dislocations $N_{\rm x}$ along the $x$-direction, while maintaining a fixed dislocation number $N_{\rm y}=4$ (with dislocation spacing parameter $n=2$) along the $y$-direction. All structures are constructed with the same number of atoms and share similar overall dimensions ($\sim118\times34\times58$~{\text{\AA}}$^3$). In contrast to MLG, which relies solely on van der Waals interactions between layers, the SDGs incorporate interlayer covalent bonds that provide additional channels for cross-plane heat transport. As anticipated, the cross-plane thermal conductivity along the dislocation axis ($z$-direction) of such structures is substantially higher by approximately an order of magnitude, than that of MLG, as shown in Fig.~\ref{fig_m_effect_TC}~(b). Moreover, the cross-plane thermal conductivity exhibits a monotonic increase with increasing dislocation number $N_{\rm x}$. To uncover the underlying mechanism of thermal transport, we analyze the real-space heat current distribution in SDGs with different $N_{\rm x}$, as depicted in Fig.~\ref{fig_m_effect_TC}~(a). The color of each atom represents the magnitude of its heat flux component along the $z$-direction, while the arrows indicate both the magnitude and direction of the per-atom heat flux. Notably, heat conduction is primarily mediated by strong covalent bonds, with only minor contributions from weak van der Waals interactions. As shown in Fig.~\ref{fig_m_effect_TC}~(c), the proportion of interlayer covalent bonds grows with increasing dislocation number $N_{\rm x}$. This enhanced network of covalent linkages leads to a corresponding increase in overall heat flux, thereby boosting cross-plane thermal transport. The observed trend clearly demonstrates that increasing the screw dislocation number $N_{\rm x}$ facilitates thermal conduction through the formation of more robust covalent interlayer pathways.

Apart from the cross-plane thermal conductivity, we also considered the effect of dislocation number $N_{\rm x}$ on the in-plane thermal conductivity along $x$-direction. Fig.~S1~(a) displays the in-plane thermal conductivity gradually increases with increasing $N_{\rm x}$. From a structural perspective, an interface exists between the dislocation region and the adjacent graphene region in SDGs. However, for the SDG structure with a larger number of dislocations, the spacing between them becomes much smaller, such that the dislocation and adjacent regions become identical, forming a unified system without a distinct interface. To gain insight into the underlying mechanism governing the impact of dislocation number on in-plane thermal conductivity, we calculate the phonon density of states (PDOS) by taking the Fourier transform of the velocity auto-correlation functions as follows,
\begin{eqnarray}
P\left(\omega\right) & = & \frac{1}{\sqrt{2\pi}}\intop_{0}^{\infty}e^{i\omega t}<\sum_{j=1}^{N}v_{j}\left(t\right)v_{j}\left(0\right)>dt, 
\label{eq_dos}
\end{eqnarray}
where $P\left(\omega\right)$ denotes PDOS at frequency $\omega$ and $v_{j}\left(t\right)v_{j}\left(0\right)$ represents the velocity auto-correlation function. The PDOS results for structures with varing $N_{\rm x}$ are shown in Fig.~\ref{fig_pdos_m_effect_TC}. Notably, for the structure with $N_{\rm x}=24$, the PDOS of region 1 and region 2 exhibit a complete overlap, indicating strong phonon coupling. In contrast, as $N_{\rm x}$ decreases, the overlaped area of PDOS in these adjacent regions is reduced, suggesting weakened phonon vibrational coupling, which contributes to reduced thermal conductivity. Overall, the in-plane thermal conductivity increases as the dislocation number $N_{\rm x}$ increases .

\subsubsection{Dislocation number $N_{\rm y}$ effect}

\begin{figure*}[tb]
  \begin{center}
    \scalebox{1.0}[1.0]{\includegraphics[width=15cm]{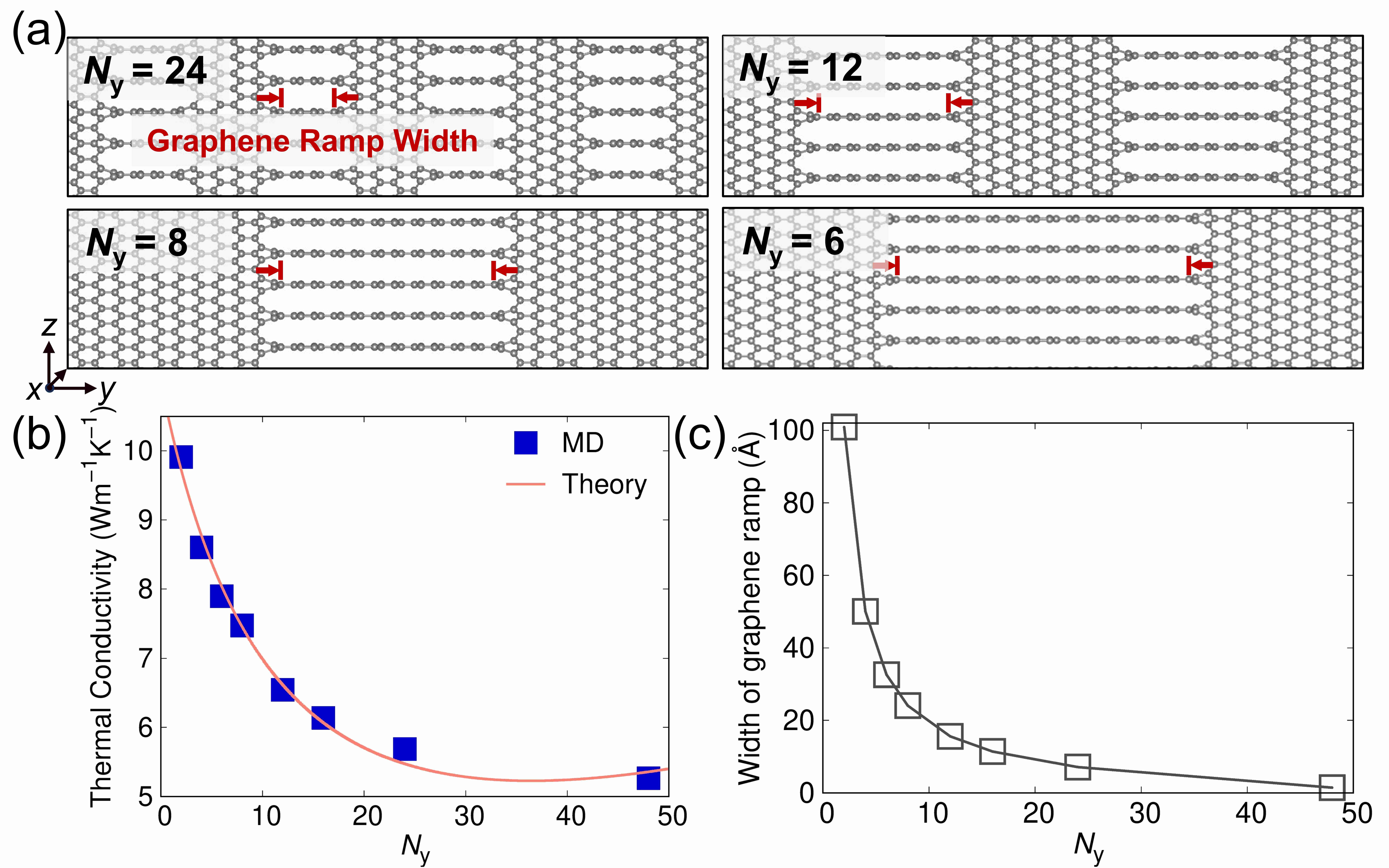}}
  \end{center}
  \caption{(Color online) The impact of screw dislocation number along $y$-direction $N_{\rm y}$ on the cross-plane thermal conductivity of SDGs. (a) Schematics of representative SDGs with different dislocation number $N_{\rm y}$, and only a small part of the system (36864 atoms in total) is shown for clarity. Variation of (b) thermal conductivity and (c) width of graphene ramp with $N_{\rm y}$. The line is the theoretical result from equation (3), respectively.} 
  \label{fig_n_effect_TC}
\end{figure*}

The number of dislocations $N_{\rm y}$ along the $y$-direction is another critical structural parameter of the SDG structure, we herein investigate its impact on thermal conductivity. Fig.~\ref{fig_n_effect_TC}~(a) illustrates a representative portion of SDGs with varying dislocation numbers $N_{\rm y}$ along the $y$-direction, while the dislocation number along the $x$-direction is fixed at $N_{\rm x}=6$ (with dislocation spacing parameter $m=2$). All structures considered have comparable dimensions of approximately $27\times204\times58$~{\text{\AA}}$^3$. As shown in Fig.~\ref{fig_n_effect_TC}~(b), the cross-plane thermal conductivity along the $z$-direction decreases with increasing $N_{\rm y}$. This trend arises because the dislocation number $N_{\rm y}$ determines the width of the graphene ramps connecting adjacent screw dislocations, and this width reduces with increasing $N_{\rm y}$, as indicated in Fig.~\ref{fig_n_effect_TC}~(c). For graphene within a narrow width, nearly all phonons, whether in-plane or flexural modes, frequently undergo lateral boundary scattering.\cite{sonvane2015length, wang2017width} As the graphene ramp narrows, the phonon mean free path limited by boundary scattering shortens, which consequently increases the rates of phonon scattering, thereby reducing the thermal conductivity. Indeed,  the heat current within the SDG structure increases as $N_{\rm y}$ decreases, which is attributable to the reduced phonon-boundary scattering. (Fig.~S2) Therefore, the cross-plane thermal conductivity decreases with increasing $N_{\rm y}$. In contrast, the in-plane thermal conductivity along the $y$-direction remains nearly constant as $N_{\rm y}$ increases, as shown in Fig.~S1~(b). Due to the presence of gaps between graphene ramps along the $y$-direction, the dislocation regions contribute negligibly to in-plane thermal transport compared to the graphene regions. As in-plane heat conduction is primarily governed by covalent bonds within the graphene domains, the variation in $N_{\rm y}$ has negligible influence on the in-plane thermal conductivity.

\subsubsection{Analytic derivation}

\begin{figure*}[tb]
  \begin{center}
    \scalebox{1.0}[1.0]{\includegraphics[width=9cm]{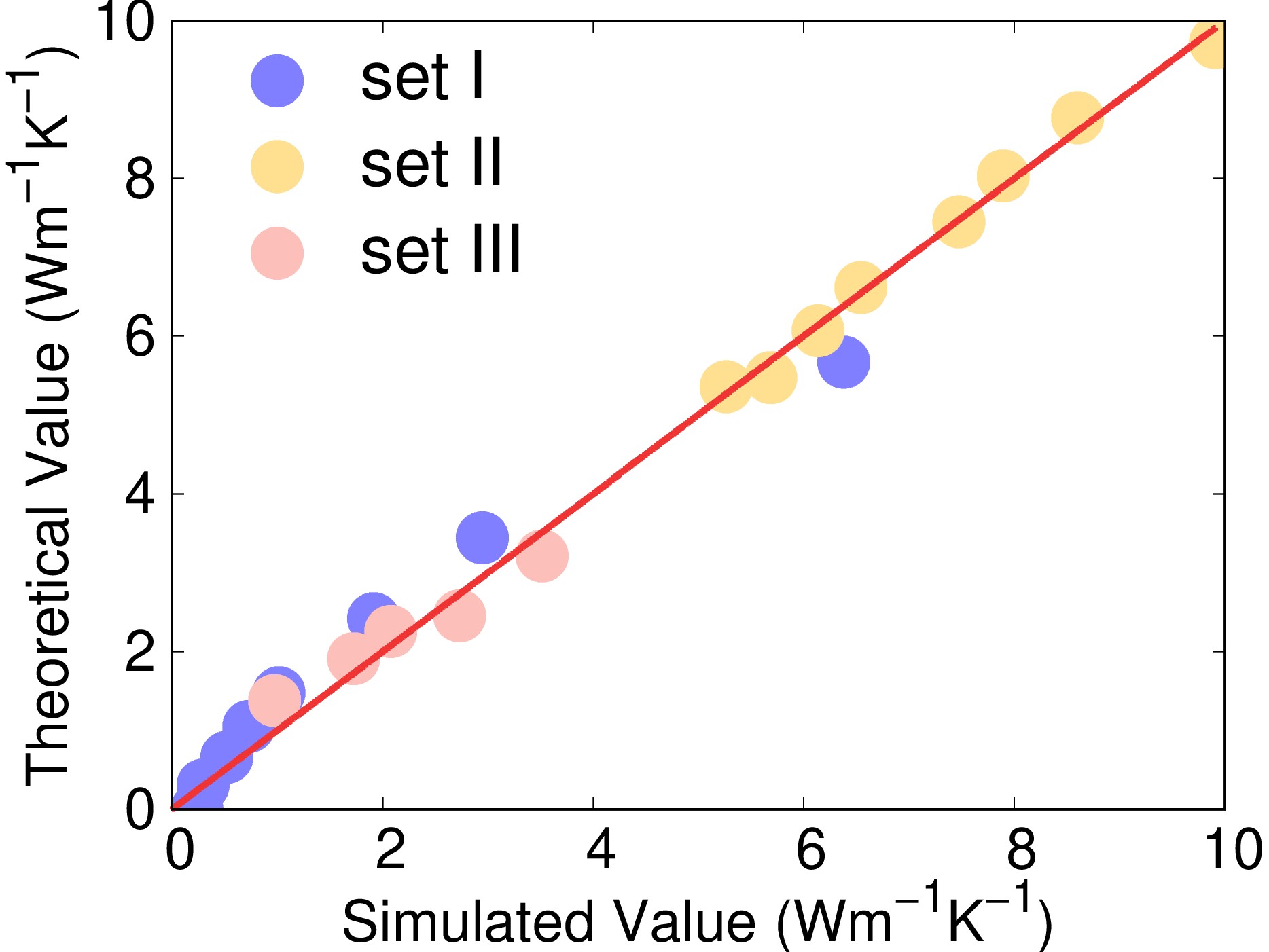}}
  \end{center}
  \caption{(Color online) Comparison between the theoretical value of the cross-plane thermal conductivity of SDGs, calculated from equation (3), and the corresponding simulation results.}
  \label{fig_verification}
\end{figure*}

Based on the above discussion, it is evident that varying the number of screw dislocations can significantly influence the cross-plane thermal conductivity of the SDG. In order to better regulate the thermal transport properties of such structure, we propose a theoretical model that explicitly incorporates the dislocation number as the key parameter. As illustrated in Fig.~\ref{fig_m_effect_TC}~(a), the structure is constructed by periodically introducing screw dislocations at specific intervals between MLG, consisting of two distinct components, dislocation regions and MLG regions. The cross-plane thermal conductivity of the SDG structure can be calculated based on the cross-plane thermal conductivity of the covalent-bonding dislocation regions and van der Waals-interacting MLG regions,
\begin{eqnarray}
\kappa & = & \alpha_{\rm bond}\kappa_{\rm bond}+(1-\alpha_{\rm bond})\kappa_{\rm vdw}\nonumber\\
 & \approx & \alpha_{\rm bond}\kappa_{\rm bond}=\frac{N_{\rm x}L_{\rm s}}{L_{\rm x}}\kappa_{\rm bond},
\label{eq_k1}
\end{eqnarray}
where $\kappa_{\rm bond}$ is the cross-plane thermal conductivity of dislocation region. Given that the cross-plane thermal conductivity of MLG region $\kappa_{\rm vdw} = 0.238$~{W/mK} is considerably lower than that of the dislocation region, the thermal conductivity contribution from the MLG is neglected. The area ratio $\alpha_{\rm bond} = A_{\rm bond}/A_{\rm tot} = N_{\rm x}L_{\rm s}/{L_{\rm x}}$ is the area of the dislocation region to that of the whole structure, where $L_{\rm x}$ is the structure size along $x$-direction and $L_{\rm s}$ is the size of one dislocation region. As shown in Fig.~\ref{fig_m_effect_TC}~(b), Eq.~(\ref{eq_k1}) captures the same trend as the simulation results, namely that the thermal conductivity increases with dislocation number $N_{\rm x}$, but a significant discrepancy exists between the theoretical and simulation results. Based on the preceding analysis of the PDOS in adjacent regions of the SDG structure, it is found that the phonon coupling between the dislocation and MLG regions strengthens as dislocation number $N_{\rm x}$ increases (Fig.~\ref{fig_pdos_m_effect_TC}), enabling heat to transfer across adjacent regions and resulting in the emergence of additional conduction pathways. As the fraction of dislocation regions increases, these additional heat transfer pathways enhance the contribution of dislocation regions to the overall thermal conductivity. Accordingly, Eq.~(\ref{eq_k1}) should be modified by a correlation coefficient $\gamma$ to account for the effect of strengthened phonon coupling,
\begin{eqnarray}
\kappa^{\ast} & = & \gamma\kappa \approx \gamma\frac{N_{\rm x}L_{\rm s}}{L_{\rm x}}\kappa_{\rm bond} ,
\label{eq_k2}
\end{eqnarray}
where correlation coefficient $\gamma=(2-e^{-\frac{N_{x}L_{s}/L_{x}}{c}})$, and $c$ is a fitting parameter. The coefficient $\gamma$ approaches 2 at high dislocation densities and approaches 1 at low dislocation densities. It should be noted that, under the condition of the maximum dislocation number $N_{x}$, no distinct interface exists between the dislocation region and the adjacent region, jointly constituting a continuous and unified system. In this case, the overall thermal conductivity is entirely dominated by the dislocation region, with its area fraction being 0.5 under such conditions, thereby constraining the maximum value of the coefficient $\gamma$ to 2.

\begin{figure*}[tb]
  \begin{center}
    \scalebox{1.0}[1.0]{\includegraphics[width=13cm]{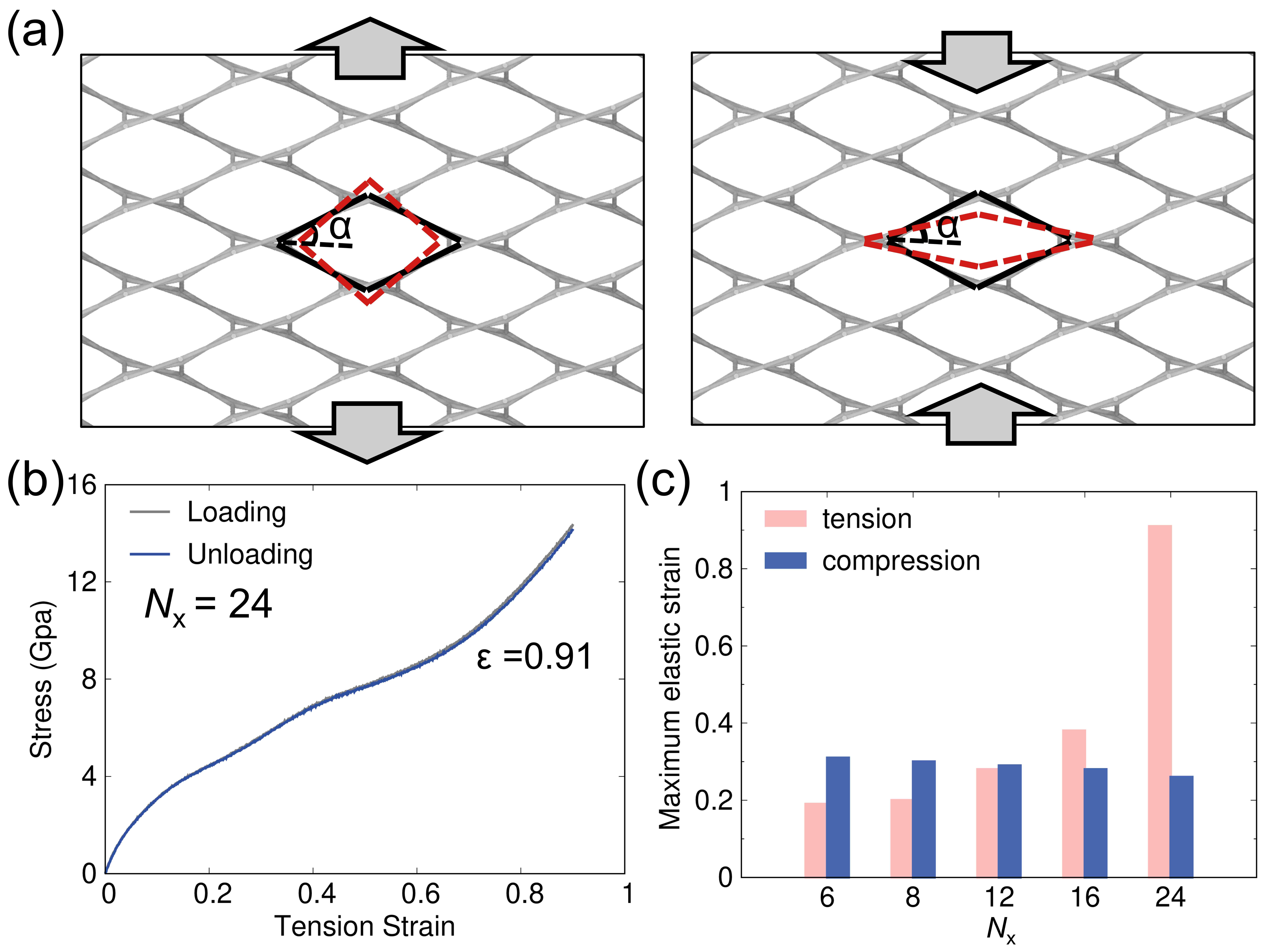}}
  \end{center}
  \caption{(Color online) Mechanical response of SDGs under tensile and compressive strain.} (a) Structural deformation of the SDG under tensile and compressive strain along the dislocation axis, with emphasis on the variation of the dislocation interface tilt angle $\alpha$. (b) Stress–strain curves of the SDG structure with $N_{\rm x}=24$ under maximum elastic tensile strain. (c) Comparison of maximum recoverable elastic strains in tension and compression for various SDGs.
  \label{fig_mechanical_test}
\end{figure*}

The dislocation region is formed by a graphene ramp bridging adjacent screw dislocations (i.e., edge) along the $y$-direction. The thermal conductivity can be decomposed into two contributions, one from the graphene ramp and the other from the edge regions,
\begin{eqnarray}
\kappa_{\rm bond} & = & \kappa_{\rm g}+\alpha_{\rm e}\kappa_{\rm e}.
\label{eq_k3}
\end{eqnarray}
Here, $\kappa_{\rm g}$ and $\kappa_{\rm e}$ are the thermal conductivities of the graphene ramp and the edge regions, respectively. The area ratio of the edge region is $\alpha_{\rm e} = \frac{2L_{\rm e}}{L_{\rm y}}N_{\rm y}$, where $L_{\rm y}$ is the structure size along the y-diretion, and $L_{\rm e}$ is the size of one edge. With increasing dislocation number $N_{\rm y}$, the dislocation region is divided into more graphene ramps, whose thermal conductivity is influenced by both boundary scattering and phonon-phonon scattering. According to Ziman’s book,\cite{ziman1960electrons} the relaxation rate of scattering can be expressed as 
\begin{eqnarray}
\frac{1}{\tau} & = & \frac{1}{\tau_{\rm bs}}+\frac{1}{\tau_{\rm pps}},
\label{eq_tau}
\end{eqnarray}
where $\tau$ is the corresponding lifetime. $\tau_{\rm bs}$ and $\tau_{\rm pps}$ are the lifetimes associated with boundary scattering and phonon-phonon scattering, respectively. The relaxation rate for boundary scattering is given by $\frac{1}{\tau_{\rm bs}}=\frac{v}{L_{\rm y}/N_{\rm y}}$, where $v$ is the phonon velocity, and $L_{\rm y}/N_{\rm y}$ is the size of the graphene ramp. The thermal conductivity contributed by the graphene ramp can be obtained from the kinetic theory,
\begin{eqnarray}
\kappa_{\rm g} & = & \frac{1}{V} \sum_{\rm q, \sigma} C_{\mathrm{ph}}(\omega) v_{\rm q,\sigma}^2 \tau_{\rm q,\sigma},
\label{eq_kg1}
\end{eqnarray}
where $V$ is the volume and $C_{\mathrm{ph}}(\omega)$ is the phonon heat capacity. We thus get the following expression from Eqs.~(\ref{eq_tau}) and (\ref{eq_kg1}),
\begin{eqnarray}
\kappa_{\rm g} & = & \frac{1}{V} \sum_{\rm q, \sigma} C_{\mathrm{ph}}(\omega) v_{\rm q,\sigma}^2\left(\frac{1}{\frac{1}{\tau_{\rm bs}}+\frac{1}{\tau_{\rm pps}}}\right)\nonumber\\
& = & \frac{1}{V} \sum_{\rm q, \sigma} C_{\mathrm{ph}}(\omega) v_{\rm q,\sigma}^2 \tau_{\rm pps}\left(\frac{1}{1+\frac{L_{\rm pps}}{L_{\rm y}}N_{\rm y}}\right)\nonumber\\
& = & \kappa_{\rm g0}\left(\frac{1}{1+\frac{L_{\rm pps}}{L_{\rm y}}N_{\rm y}}\right).
\label{eq_kg2}
\end{eqnarray}
Here, $\kappa_{\rm g0}$ denotes the intrinsic thermal conductivity of the graphene ramp in the absence of boundary scattering, and $L_{\rm pps}$ is the mean free path associated with phonon-phonon scattering. An analytic expression can be achieved for the thermal conductivity of the covalent-bonding dislocation region by combining (\ref{eq_k3}) and (\ref{eq_kg2}),
\begin{eqnarray}
\kappa_{\rm bond} & = & \kappa_{\rm g0}\left(\frac{1}{1+\frac{L_{\rm pps}}{L_{\rm y}}N_{\rm y}}\right)+\kappa_{\rm e}\frac{2L_{\rm e}}{L_{\rm y}}N_{\rm y}.
\label{eq_k4}
\end{eqnarray}
Fig.~\ref{fig_n_effect_TC}~(b) shows that the analytic result in Eq.~(\ref{eq_k2})  is in good agreement with the MD simulation results for SDGs with different dislocation numbers. Here, the quantity $L_{\rm pps}N_{\rm y}/L_{\rm y} = L_{\rm pps}/L_{\rm bs}$ represents the ratio of the phonon–phonon scattering mean free path to that associated with boundary scattering. Our numerical calculations yield a fitting value of $L_{\rm pps}/L_{\rm y} = 0.07$.  Accordingly, when the dislocation number $N_{\rm y}$ is relatively small, phonon–phonon scattering dominates over boundary scattering, as reflected by $L_{\rm pps}/L_{\rm bs}<1$. In contrast, when $N_{\rm y}$ becomes larger, boundary scattering becomes the dominant mechanism, as indicated by $L_{\rm pps}/L_{\rm bs}>1$. These results are reasonable, as tuning the number of dislocations modifies the dimensions of the graphene ramps, leading to a competitive relationship between boundary scattering and phonon–phonon scattering. We obtain numerically these parameters, where the intrinsic thermal conductivity $\kappa_{\rm g0} = 13.1$~{W/mK} is close to that of the structure containing only two edges, and thermal conductivity of the edge region $\kappa_{\rm e} = 3.4$~{W/mK} is lower than $\kappa_{\rm g0}$. 

Finally, we compare the numerical results with the predictions of the derived theoretical formula to evaluate its applicability (Fig.~\ref{fig_verification}). In addition to the two simulation sets (Set I and Set II) that investigate the effects of dislocation numbers $N_{\rm x}$ and $N_{\rm y}$, a third set (Set III) is carried out by varying both the structure size and the dislocation numbers. In Set III, the model size is fixed at $59\times102\times 58$~{\text{\AA}}$^3$, with dislocation numbers $(N_{\rm x}, N_{\rm y})$ of (3, 3), (3, 6), (4, 4), (4, 6) and (6, 6). As shown in Fig.~\ref{fig_verification}, the numerical values agree well with the theoretical values, confirming the validity of the proposed theoretical formula. On this basis, this theoretical formula can serve as a practical guideline for designing SDGs with tailored thermal conductivities, either high or low, depending on application demands.

\subsection{Anomalous strain-dependent thermal conductivity}

\begin{figure*}[tb]
  \begin{center}
    \scalebox{1.0}[1.0]{\includegraphics[width=15cm]{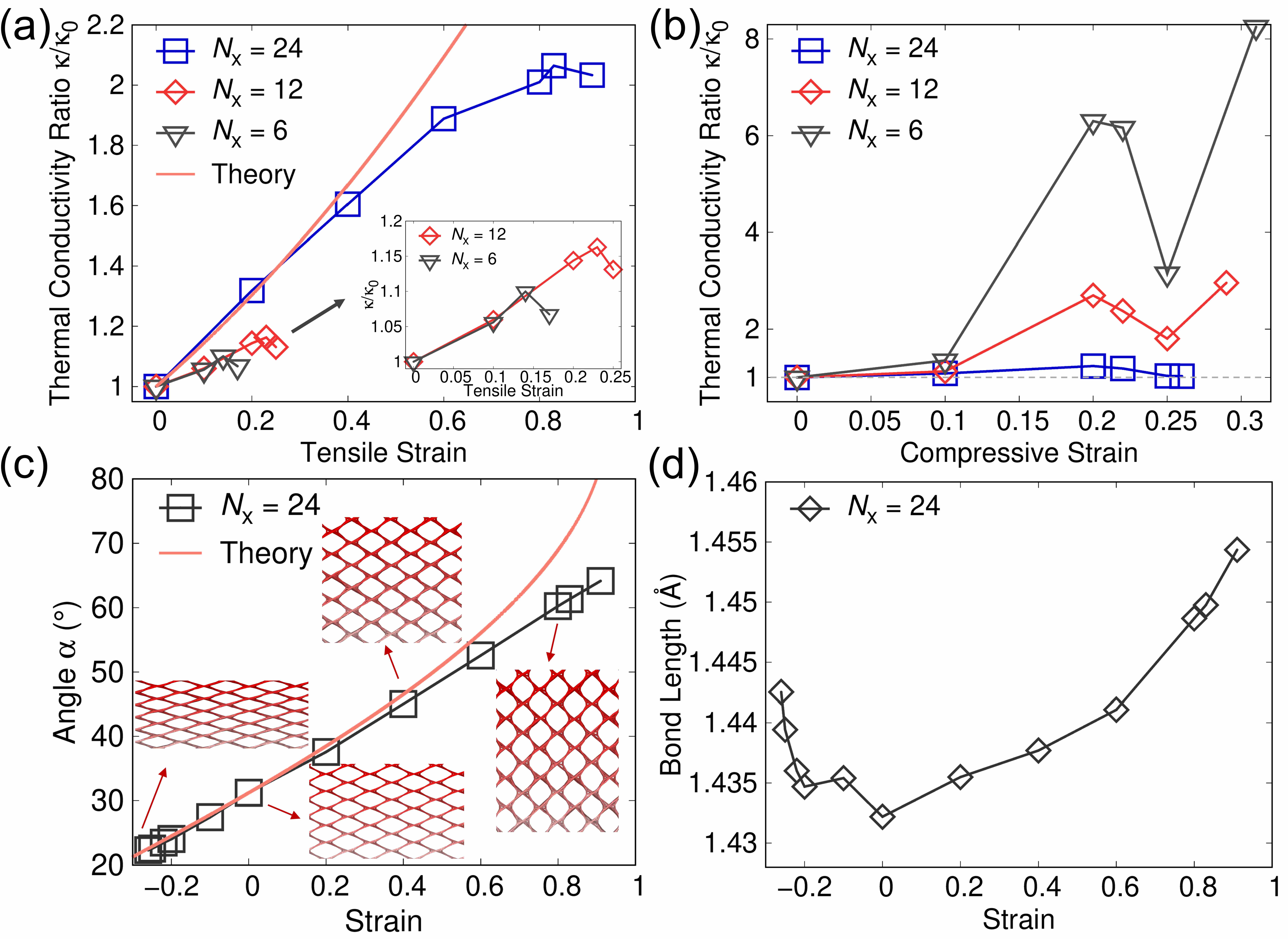}}
  \end{center}
  \caption{(Color online)  Strain-induced variation of cross-plane thermal conductivity in SDGs. Variation of relative cross-plane thermal conductivity (the ratio of thermal conductivity under strain, $\kappa$, to that of the unstrained case, $\kappa_{0}$) of various SDGs under elastic (a) tensile and (b) compressive strain along dislocation axis. The line is the theoretical result from equation (12). (c) Dislocation angle in the SDG structure with $N_{\rm x}=24$ as a function of strain. The insets in (c) show the snapshots of the structure at zero strain, 26\% compressive strain, and 40\% and 80\% tensile strains, respectively. The line is the theoretical result from equations (10). (d) Strain dependence of the average bond length within the dislocation interface. } 
  \label{fig_strain_effect_TC}
\end{figure*}

\subsubsection{Superelasticity of SDGs}

Strain has long been recognized as an effective strategy for modulating thermal conductivity of materials. In this section, we examine how mechanical strain influences the thermal transport behavior of SDGs. We first test the mechanical stability of a series of SDGs with different dislocation number. As depicted in Fig.~\ref{fig_mechanical_test}~(a), tensile and compressive simulations are performed along the dislocation axis, accompanied by a variation in the angle between graphene ramp connecting two adjacent dislocations and and $x$-direction, referred to as the dislocation interface tilt angle $\alpha$. Since changes in the dislocation number $N_{\rm x}$ affect the number of interlayer covalent bonds between graphene layers, this has a critical impact on the mechanical stability and thermal transport along the dislocation axis, in contrast to pristine MLG. Accordingly, we select a representative set of SDGs with varying dislocation number $N_{\rm x}$ and fixed dislocation number $N_{\rm y}=2$ to investigate the mechanical and thermal responses. 

Under tensile loading along the dislocation axis, the SDG structure with the maximum dislocation number $N_{\rm x}=24$ (within the considered dimensional range) can sustain tensile strains up to 100\%, as illustrated in Fig.~S3~(a). Beyond this threshold, partial bond breaking occurs in the dislocation region, relieving a portion of the applied stress. To further assess the elastic limit, loading–unloading simulations are performed to determine the maximum recoverable elastic tensile strain. As shown in Fig.~\ref{fig_mechanical_test}~(b), the SDG structure with $N_{\rm x}=24$ exhibits an elastic tensile strain of up to 91\%. This enhanced elasticity is attributed to the increasing number of covalent bonds bridging graphene layers. As displayed in Fig.~\ref{fig_mechanical_test}~(c), the maximum elastic tensile strain sustainable by the SDG structure increases with the dislocation number $N_{\rm x}$. The SDG structure with $N_{\rm x} = 24$ does not exhibit structural failure under 70\% strain even at temperatures as high as 1000~{K} (Fig.~S4~(a)), indicating its ability to withstand large deformations at high temperature. Moreover, under a maximum elastic tensile strain of 91\%, this structure remains thermodynamically stable at room temperature (Fig.~S4~(b)). Similarly, other structures with various dislocation numbers also retain structural stability under maximum tensile strain at temperatures up to 1000~{K}. On the other hand, the compressive behavior of these structures remains nearly consistent across different $N_{\rm x}$ values, with a slight decline in compressive strength as $N_{\rm x}$ increases. The stress–strain response in Fig.~S5~(a) reveals a two-stage behavior, in the initial stage, stress decreases due to partial bond breaking at the dislocation interface; in the subsequent stage, the stress rises sharply, driven by the formation of a large number of diamond-like sp³-hybridized structures. Across all tested configurations, the maximum recoverable elastic compressive strain is approximately 30\%.

\subsubsection{Unusual thermal conductivity}

The superelasticity of SDGs enables broad strain-induced modulation of thermal conductivity. We next analyze this strain–conductivity relationship in detail. Fig.~\ref{fig_strain_effect_TC}~(a) illustrates the change in cross-plane thermal conductivity of SDGs with different dislocation number $N_{\rm x}$ throughout the full recoverable elastic tensile strain range along dislocation axis, relative to the unstrained condition. Notably, the cross-plane thermal conductivity exhibits a significant increase with applied tensile strain. More specifically, the cross-plane thermal conductivity of the SDG structure with $N_{\rm x}=24$ increases by more than 100\% under above 80\% tensile strain. This unusual increase in thermal conductivity of the SDG structure under tensile strain contrasts sharply with the conventional one-dimensional screw-dislocation carbon structure, wherein tensile-induced phonon modes softening reduces phonon group velocities and consequently lowers thermal conductivity.\cite{zhan2018graphene} Only few studies on enhanced thermal conductivity in 3D carbon architectures under tensile strain have been reported, and the achievable tunability of thermal transport remains generally limited, typically confined within the range of 10\% tensile strain.\cite{kuang2015unusual, han2018unusual, wang2022anomalous} In contrast, the 3D SDGs, featuring exceptional superelasticity, enables broad thermal conductivity modulation over an ultra-large, fully recoverable elastic strain range without triggering any phase transitions.

We herein take the structure with $N_{\rm x}=24$ as a representative example to investigate the fundamental mechanism by which strain affects heat transport in the SDGs. A key structural feature for the SDG structure under strain is the change in the dislocation interface tilt angle $\alpha$ (Fig.~\ref{fig_mechanical_test}~(a)). We analyze the evolution of angle $\alpha$ across the entire range of recoverable elastic compression and tension for the SDG structure with $N_{\rm x}=24$. As shown in Fig.~\ref{fig_strain_effect_TC}~(c), the angle $\alpha$ increases monotonically with applied strain, accompanied by corresponding structural changes. This change extends the propagation path of heat flow along the dislocation axis, thereby enhancing thermal transport in the cross-plane direction. Considering the effect of stretching on the interatomic interaction strength, we also examine the changes in covalent bond lengths on the dislocation region as a function of strain, shown in Fig.~\ref{fig_strain_effect_TC}~(d). When the applied tensile strain is below 60\%, the bond lengths increase slowly and linearly by approximately 0.6\%. Once the strain beyond this point, the covalent bonds elongate more rapidly, and the thermal conductivity begins to decrease due to the softened phonon modes and increased lattice anharmonicities.\cite{li2010strain} The negative impact of bond lengthening competes with the positive contribution from increased dislocation interface tilt angle. As a consequence, the unusual enhancement in cross-plane thermal conductivity of SDGs under tensile strain mainly results from the elongation of heat transfer path along the dislocation axis, which increases cross-plane heat flux. Furthermore, this unusual strain-dependent thermal conductivity is also expected in SDGs with different dislocation numbers $N_{\rm x}$ (Fig.~\ref{fig_strain_effect_TC}~(a)).

The enhancement of thermal conductivity under tensile strain is an unusual phenomenon, especially given the broad tunable range observed. To gain and verify the underlying modulation mechanism, we extend the analytical model developed in the last section to account for only simple tensile strain effects. Assuming that bond lengths remain unchanged under the applied strain, a relationship between the dislocation interface tilt angle $\alpha$ and the strain $\varepsilon$ can be established as follows,
\begin{eqnarray}
\varepsilon & = & \frac{\sin\alpha-\sin\alpha_{0}}{\sin\alpha_{0}},
\label{eq_strain}
\end{eqnarray}
where $\alpha_{0}$ is the dislocation interface tilt angle under the unstrained condition. From Eq.~(\ref{eq_strain}), we can obtain an explicit expression for the angle $\alpha$,
\begin{eqnarray}
\alpha & = & \arcsin((\varepsilon+1)\sin\alpha_{0}).
\label{eq_angle}
\end{eqnarray}
As illustrated in Fig.~\ref{fig_strain_effect_TC}~(c), the theoretical values of the angle $\alpha$ agree well with the simulation results. The deviation observed under large strains may arise from the assumption of invariant bond lengths, which in reality are elongated under high tensile strain. The applied tensile strain increases the angle $\alpha$, thereby altering the dimensions of the dislocation region along both the $x$- and $z$-directions, ${L_{\rm z}}/{L_{\rm x}}= \tan\alpha$. This structural change increases heat flux along the dislocation axis, i.e., the $z$-direction. According to Fourier’s law, the thermal conductivity can be expressed as follows,
\begin{eqnarray}
\kappa & = & \frac{Q}{A\nabla T}=\frac{Q}{L_{\rm y}L_{\rm x}\frac{\triangle T}{L_{\rm z}}}\nonumber\\
 & = & \kappa_{0}\frac{\tan\alpha}{\tan\alpha_{0}}\nonumber\\
 & = & \kappa_{0}\cdot\cos\alpha_{0}\cdot\frac{\left(\varepsilon+1\right)}{\sqrt{(1-\left(\varepsilon+1\right)^{2}\sin^{2}\alpha_{0})}},
\label{eq_kstrain1}
\end{eqnarray}
where $Q$ denotes the thermal energy transferred across a section per unit time. $A = L_{\rm x}\times L_{\rm y}$ represents the cross-sectional area perpendicular to the direction of heat flow, $L_{\rm z}$ is the heat conduction length, $L_{\rm x}$ and $L_{\rm y}$ are the length of the sample in the $x$- and $y$-directions. $\nabla T$ is the temperature gradient, and $\triangle T$ is the temperature difference across the sample. $\kappa_{0}$ is the thermal conductivity under the unstrained condition. A second-order Taylor approximation yields the following analytical expression for thermal conductivity as a function of tensile strain,
\begin{eqnarray}
\kappa & \approx & \kappa_{0}(1+c_{1}\cdot\varepsilon+c_{2}\cdot\varepsilon^{2}),
\label{eq_kstrain2}
\end{eqnarray}
where parameter $c_{1}=\frac{1}{\cos^{2}\alpha_{0}}=1.368$, and parameter $c_{2}=\frac{3\sin^{2}\alpha_{0}}{2\cos^{4}\alpha_{0}}=0.755$ are dimensionless. Fig.~\ref{fig_strain_effect_TC}~(a) shows good agreement between the theoretical predictions and simulation results. The deviation observed at high strain levels is attributed to the stretching of atomic covalent bonds, which softens the phonon modes and consequently reduces the thermal conductivity.

Like tensile strain, compressive strain also enhances the cross-plane thermal conductivity of SDGs, as shown in Fig.~\ref{fig_strain_effect_TC}~(b). The key difference is that compressive strain strengthens van der Waals interactions, boosting phonon coupling and group velocities, thereby improving thermal transport efficiency.\cite{ding2015plane, meng2019thermal, yang2024tuning} However, once the applied compressive strain exceeds 20\%, a decline in thermal conductivity is observed; this trend reverses when the strain surpasses 25\%. No structural damage is observed, but the radial distribution function reveals increased medium-range disorder under 20–25\% strain (see Fig.~S6), likely causing stronger phonon scattering and thus reducing thermal conductivity. With further increase in compressive strain, the structure regains medium-range order, and interatomic interactions are strengthened, resulting in a renewed increase in thermal conductivity. At approximately 30\% compressive strain, the cross-plane thermal conductivity of structure with $N_{\rm x} = 6$ exhibits a remarkable enhancement of over 700\%. The regulation capability under compressive strain of SDGs surpasses that of most 3D carbon allotropes. For comparision, the thermal conductivity of diamond increases by only four times even under extreme stress of 400~{GPa}.\cite{broido2012thermal} In contrast, SDGs with a greater number of dislocations display minimal changes in thermal conductivity under compression. This discrepancy stems their distinct heat transfer mechanisms. In SDGs with more dislocations, thermal transport is primarily dominated by covalent bonds. Applied compressive strain reduces the angle $\alpha$ and shortens the heat transfer path (see Fig.~\ref{fig_strain_effect_TC}~(c)), thereby suppressing heat flux along the dislocation axis. This negative effect largely offsets the positive contribution from enhanced van der Waals interactions. Conversely, in SDGs with fewer dislocations where van der Waals interactions dominate, the strengthened interactions under compression lead to substantial increase in thermal conductivity. 

\section{Conclusion}

In summary, our MD simulations and thermodynamic analyses reveal that the 3D topological carbon allotropes, SDGs, exhibit an unusual increase in cross-plane thermal conductivity when subjected to tensile strain along the dislocation axis. This counterintuitive enhancement is attributed to a strain-induced increase in the dislocation interface tilt angle, which enlongate heat transport path along the dislocation axis. Notably, such structure can exhibit an anomalous enhancement of more than 100\% in cross-plane thermal conductivity under tensile strains up to 80\%. Additionally, compressive strains also lead to a substantial increase in the cross-plane thermal conductivity, with an increase of over 700\% under compressive strains up to 30\%. By tuning the dislocation numbers, a variety of structurally distinct SDGs can be achieved. We also systematically analyze the dependence of dislocation number on the thermal behavior, and derive an analytical model that correlates dislocation number and strain with cross-plane thermal conductivity, providing a theoretical foundation for the targeted design and modulation of thermally efficient screw-dislocated carbon materials. These findings unveil a class of 3D topological carbon allotropes that combine robust topological electronic states with strain-stable, or even strain-enhanced, thermal transport properties, offering promising opportunities for improving the performance of advanced flexible and wearable electronic devices.

\section{Methods}

The thermal conductivity $\kappa$ is calculated using the NEMD method based on Fourier\textquotesingle s  law,\cite{arfken2012international} 
\begin{eqnarray}
J & = & -\kappa \frac{dT}{dx},
\label{eq_J}
\end{eqnarray}
where the heat flux $J$ represents the a flow of energy per unit area per unit time and $\frac{dT}{dx}$ denotes the temperature gradient along the heat flow direction. During the thermal conductivity calculation, the regions at both ends of the heat transfer direction are fixed as illustrated Fig.~\ref{fig_model}~(c). The heat source and  heat sink are maintained at a temperature difference of 20~{K} from the average temperature. We analyze the temperature gradient and heat flux density to obtain the thermal conductivity. 

In the elastic tension and compression simulations, the structure is stretched (or compressed) along the dislocation axis at a specific
strain rate of 0.1\% per picosecond. Once a certain strain is reached, loading is applied to the structure in the opposite direction at the same rate to achieve unloading. Throughout the simulation, the stress along the non-loading direction is controlled at zero within Berendsen barostat. By recording the stress-strain data and observing whether the loading and unloading curves coincide, we determine the structure\textquotesingle s maximum recoverable elastic stretching and compressive strains.

We employ NEP-C\cite{li2023vacancy}, a machine learning potential optimized for carbon systems, to calculate thermal transport properties with high accuracy and low cost. NEP-C is trained on the GAP-20 dataset,\cite{rowe2020accurate} which encompasses a broad spectrum of carbon allotropes and defect structures (e.g., vacancy, Stone-Wales). Taking the SDG structure with dislocation spacing parameter $m=2$ and $n=1$ as an example, a comparison of the phonon dispersion curves calculated using NEP-C and DFT shows good agreement as presented in Fig.~S7. It is clearly that the NEP-C potential is sufficient for studying the thermal properties of SDGs. Simulations are performed using the GPUMD package\cite{fan2022gpumd}, and visualizations are carried out with the Open Visualization Tool (OVITO)\cite{stukowski2009visualization}. The standard Newton equations of motion are integrated in time using the velocity Verlet algorithm with a time step of 0.5~{fs}. 

\textbf{Acknowledgment} This work is supported by the National Natural Science Foundation of China (Grant Nos. 12072182 and 12421002) and the National Research Foundation, Singapore under Award No. NRF-CRP24-2020-0002.

\bibliographystyle{nature}
\bibliography{biball}

\end{document}